\documentclass[usenatbib]{mn2e}
\usepackage{graphicx}
\usepackage{txfonts}
\usepackage{threeparttable}
\usepackage{aas_macros}
\usepackage{color}

\title[Faint progenitors of luminous $z \sim 6$ quasars: why don't we see them?]{Faint progenitors of luminous $z \sim 6$ quasars: why don't we see them?}
\author[Pezzulli et al.]{Edwige Pezzulli$^{1,2,3}$\thanks{E-mail:
edwige.pezzulli@oa-roma.inaf.it}, Rosa Valiante$^{2}$, Maria C.  Orofino$^{4}$, Raffaella Schneider$^{1}$, \and Simona Gallerani$^{4}$ and Tullia Sbarrato$^5$ \\
$^{1}$Dipartimento di Fisica, Universit{\'a} di Roma  ``La Sapienza'', P.le Aldo Moro 2, 00185, Roma, Italy \\
$^{2}$INAF/Osservatorio Astronomico di Roma, Via di Frascati 33, 00040 Monte Porzio Catone, Italy\\
$^{3}$INFN, Sezione di Roma I, P.le Aldo Moro 2, 00185 Roma, Italy \\
$^{4}$Scuola Normale Superiore, Piazza dei Cavalieri 7, 56126, Pisa, Italy\\
$^{5}$Dipartimento di Fisica "G. Occhialini", Universita di Milano - Bicocca, Piazza della Scienza 3, 20126 Milano, Italy}
\begin{document}

\date{11 December 2016}

\pagerange{\pageref{firstpage}--\pageref{lastpage}} \pubyear{2016}

\maketitle
\label{firstpage}

\begin{abstract}
Observational searches for faint active nuclei at $z > 6$ have been extremely
elusive, with a few candidates whose high-$z$ nature is still to be confirmed. 
Interpreting this lack of detections is crucial to improve our understanding of high-$z$ supermassive black holes (SMBHs) formation and growth. In this work, we present a model for the emission of accreting BHs in the X-ray band, taking into account super-Eddington accretion, which can be very common in gas-rich systems at high-$z$. We compute the spectral energy distribution for a sample of active galaxies simulated in a cosmological context, which represent the progenitors of a $z \sim 6$ SMBH with $M_{\rm BH} \sim 10^9 \, M_\odot$. We find an average Compton thick fraction of $\sim 45\%$ and large typical column densities ($N_H \gtrsim 10^{23} \rm \, cm^2$). However, faint progenitors are still luminous enough to be detected in the X-ray band of current surveys. Even accounting for a maximum obscuration effect, the number of detectable BHs is reduced at most by a factor 2. In our simulated sample, observations of faint quasars are mainly limited by their very low active fraction ($f_{\rm act} \sim 1 \%$), which is the result of short, super-critical growth episodes. We suggest that to detect high-$z$ SMBHs progenitors, large area surveys with shallower sensitivities, such as \textit{Cosmos Legacy} and XMM-LSS+XXL, are to be preferred with respect to deep surveys probing smaller fields, such as CDF-S.
\end{abstract}



\section{Introduction}
The nature of the first supermassive black holes (SMBHs), powering the most luminous quasars observed at $z\sim 6$, is still far from being understood. These actively accreting BHs of $10^9-10^{10}$ M$_\odot$ must have formed and grown in less than 1 Gyr.

The Eddington luminosity $L_{\rm Edd}$, defined as the maximum luminosity 
that a black hole (BH) can achieve, as a result of the balance between radiation and gravitation,
classically provides a limit to the rate at which a BH can accrete gas.

If we assume that the BH accretes a fraction
$(1 - \epsilon_r)$ of the infalling material, at the Eddington rate
$\dot{M}_{\rm Edd,1} = L_{\rm Edd}/c^2$, its mass growth can
be described as

\begin{equation}\label{eq:massevo}
	M_{\rm BH}(t) = M_{0} e^{\frac{1 - \epsilon_r}{\epsilon_r}\frac{t}{t_{\rm \small{Edd}}}},
\end{equation}

\noindent where $\epsilon_r$ is the radiative efficiency, $M_{0}$ is the initial mass of the seed BH and $t_{\rm Edd} \sim 0.45$ Gyr is the Eddington time. Two main seed formation mechanisms have been proposed (see e.g. \citealt{Volonteri2008,Volonteri2009,Volonteri2010,Volonteri2012} and \citealt{Latif2016} for a review).
One scenario predicts {\it light seeds} of $M_0 \sim 100 \, M_\odot$,
consisting of Population III (Pop~III) stellar remnants (\citealt{Madau2001,Volonteri2003}). The second model predicts a higher seed mass,
formed via the {\it direct collapse} of gas onto $M_0 \simeq [10^4 - 10^6] \, M_\odot$ BH (\citealt{Haehnelt1993, Bromm2003, Begelman2006, Lodato2006}).

The Eddington limit provides a tight constraint on the value of $M_0$.
To reproduce the mass of ULAS J1120, $M_{\rm SMBH} \sim 2 \times 10^9 \, M_\odot$, the most distant quasar currently known at $z \sim 7$ \citep{Mortlock2011}, the initial seed has to be $M_0 \gtrsim 4 \times 10^3 \, M_{\odot}$ if $\epsilon_r \sim 0.1$ and the BH has accreted uninterruptedly since $z = 30$\footnote{Hereafter we adopt a Lambda Cold Dark Matter ($\Lambda$CDM) cosmology 
with parameters $\Omega_{\rm M} = 0.314$, $\Omega_{\Lambda} = 0.686$, and $h = 0.674$ (Planck Collaboration et al. 2014).} at the Eddington rate.

The assumption of such uninterrupted mass accretion is unrealistic. In fact, the accretion rate is
limited by the available gas mass and by the radiative feedback produced by the accretion process itself.
An alternative possibility is to have short, episodic periods of super-Eddington accretion,
that allow to grow a SMBH mass even starting from light seeds \citep{Haiman2004, Yoo2004, Shapiro2005, Volonteri2005, Pelupessy2007, Tanaka2009, Madau2014, Volonteri2015, Lupi2016, Pezzulli2016}.

The detection and characterization of $z>6$ quasars fainter than the ones currently observed would be extremely helpful to improve our understanding of the high-z SMBHs formation process.
Several observational campaigns in the X-ray band have been made to discover the faint progenitors of SMBHs at $z \gtrsim 5$. 
\citet{Weigel2015} searched for active galactic nuclei (AGNs) in the \textit{Chandra} Deep Field South (CDF-S) starting their analysis from already X-ray selected sources within the \textit{Chandra} 4 Ms catalogue \citep{Xue2011}. They combined GOODS, CANDELS and \textit{Spitzer} data to estimate the photometric redshift of their sources but no convincing AGN candidates was found at $z \gtrsim 5$.  This
result has been confirmed by the independent analysis of \citet{Georgakakis2015}, who combined deep \textit{Chandra} and wide-area/shallow XMM–-Newton survey fields to infer the evolution of the 
X-ray luminosity function at $3 \lesssim z \lesssim 5$. They find a strong evolution at the faint-end and extrapolating this trend to $z \gtrsim 5$ they predict $< 1$ AGN in the CDF-S.
A complementary approach was followed by \citet{Treister2013}, who started from a sample of photometrically selected galaxies at $z \sim 6$, $7$, and $8$ from the {\it Hubble Space Telescope} Ultra Deep
Field (HUDF) and CANDELS, and then combined these data with the 4 Ms CDF-S. None of the sources was detected in X-ray either individually or via stacking, placing tight constraints on black hole growth at these redshifts\footnote{These authors estimate an accreted mass density $ \rm < 1000 \, M_{\odot}\, Mpc^{-3}$ at $z \sim 6$.}.
More recently, \citet{Vito2016} investigated the X-ray emission of samples of CANDELS selected galaxies at redshift $3.5 \leq z \leq 6.5$, stacking the data from 7 Ms CDF-S. Assuming that all the X-ray stacked emission is due to X-ray binaries, the authors find that their inferred star formation rate density is consistent with the UV-based result in the literature. This suggests that most of the X-ray emission from individually undetected galaxies is due to binaries.

However, by improving the multi-dimensional source detection technique developed by \citet{Fiore2012}, \citet{Giallongo2015} identified three faint AGN candidates in the GOODS-S field, with photometric redshifts $z>6$. Very faint $z>4$ galaxies are selected in the sample from the near infrared (NIR) H band luminosity, down to $H\leq 27$ (which at these redshifts corresponds to a UV rest-frame selection). Then, AGN candidates with soft X-ray ($[0.5-2]$ KeV) fluxes above $F_{\rm X}\sim 1.5\times 10^{-17} \rm erg \, s^{-1}\,cm^{-2}$, are extracted from the sub-sample. NIR-based selection methods allow to reach fainter X-ray fluxes than direct blind X-ray selections. 
By means of a novel photometric method, supported by numerical simulations, \citet{Pacucci2016} identified two of these high redshift AGN candidates, object 33160 at $z\sim 6$ and object 29323 at $z\sim 9.7$, as possible hosts of direct collapse BHs.

In contrast, none of the $z>6$ NIR-selected sources identified by \citet{Giallongo2015} are found by \citet{Cappelluti16} in the same area, using a similar approach as in \citet{Giallongo2015} but different thresholds and energy bands. 
Beside the poor statistics and the large uncertainties related to photometric redshift estimates\footnote{An example is the source 29323 with the highest photo-z=9.7 selected by \citet{Giallongo2015} but excluded from the \citet{Cappelluti16} sample because of artifacts in the spectral energy distribution.},   
the authors underline that the actual number of high redshift AGN candidates is very sensitive to the adopted selection procedure. 
The analysis of future surveys carried out with the next generation X-ray observatory \textit{ATHENA+}, will enlarge the systematic search of high redshift AGNs to lower luminosity sources.

Possible explanations to the very limited number (or even the lack) of $z>6$ detections reported in these studies, are strong gas and dust obscuration \citep{Gilli2007, Treister2009, Fiore2009} or low BH occupation fraction (i.e. a low fraction of halos containing a BH in their centres). 
For this reason, several authors have proposed to search for SMBH progenitors through far-infrared emission lines that are unaffected by dust obscuration (e.g. \citealt{Spaans2008}, \citealt{Schleicher2010}, \citealt{Gallerani2014}).
Additionally, short episodes of mildly super-Eddington growth, followed by longer periods of quiescence, with duty cycles of $20-50\%$ \citep{Madau2014}, may further decrease the probability of observing accreting BHs, resulting in a low \textit{active} BH occupation fraction. It should be noted that BHs cannot be detected by X-ray observations if their growth is driven by BH-BH mergers, rather than mass accretion. Indeed, 
the accretion process is directly related to the emission in this band (see the detailed discussion by \citealt{Treister2013}).

In this work, we want to understand which of these explanations is the most plausible to interpret the shortage of detections of high-$z$ faint BHs. 
To this aim, we investigate the detectability of progenitors of $z\sim 6$ SMBHs in the super-critical growth scenario, by constructing a model for the optical/UV and X-ray emission of an active BH. We consider the dependence of the X-ray spectrum on the Eddington ratio $\lambda_{\rm Edd} = L_{\rm bol}/L_{\rm Edd}$ (i.e. the bolometric-to-Eddington luminosity ratio). We apply the emission model to the sample of $z > 6$ BH progenitors of $z \sim 6$ quasars analysed in \citet[][hereafter P16]{Pezzulli2016}.
The sample has been generated using the data-constrained semi-analytical model \textsc{GAMETE/QSOdust}, that allows to simulate a statistically meaningful number of hierarchical histories of $z \sim 6$ quasars, following the star formation history, chemical evolution and nuclear black hole growth in all their progenitor galaxies. The model has been thoroughly described in \citet{Valiante2011,Valiante2012,Valiante2014} and P16.

In P16, we analysed the importance of super-Eddington accretion for the formation of $z \sim 6$ quasars assuming that Pop~III BH remnants of $\sim 100 \, M_\odot$ grow via radiatively inefficient \textit{slim} accretion discs \citep{Abramowicz1988}. We found that $\sim 80\%$ of the final SMBH mass grows via super-critical episodes, that represent the most widespread accretion regime down to $z \sim 10$.  Moreover, rapid accretion in dense, gas-rich environments allows to grow, on average, a BH mass of $10^4 M_{\odot}$ at $z \sim 20$, comparable to that of direct collapse BHs.

The paper is organized as follows: in Section \ref{Sec1} we describe the developed model for the spectrum of accreting BHs, in Section \ref{sample} we analyse the properties of the simulated BH sample, while in Section \ref{results} we present our results for the observability of faint SMBHs progenitors with current and future surveys. Finally, conclusions are drawn in Section \ref{conclusions}.

\section{The Spectral Energy Distribution of accreting BHs}\label{Sec1}

The spectral energy distribution (SED) of AGNs has been modelled in the literature using empirical models inferred from  observations (e.g. \citealt{Marconi2004,Lusso2010}) or calibrating physically motivated prescriptions with observations \citep{Yue2013}.
These models have been also applied, when necessary, to super-critical growth regimes \citep{Pacucci2015}. 
Simulations of \textit{slim} discs have been also developed, taking into account the vertical disc structure and predicting the SED of the emitted radiation \citep{Wang1999,Watarai2000,Ohsuga2003,Shimura2003}.

The typical spectrum of a radio quiet AGN can be approximately divided into three major components: the Infrared Bump (IB), the Big Blue Bump (BBB), and the X-ray region.
Under the assumption of an optically thick disc, a large fraction, up to $\gtrsim 50 \%$, of the bolometric emission is expected to be in the form of optical/UV thermal disc photons, producing the BBB continuum that extends from the NIR at $1 \mu m$ to the UV $\sim$ 1000 $\AA$ or the soft X-ray wavelengths, in some cases.
In the hard X-ray band the AGN flux per unit frequency $F_{\nu}$ is well described by a power law with spectral index $\sim 0.9$ \citep{Piconcelli2005,Just2007}. This emission is due to Compton up-scattering of optical/UV photons by hot electrons in the corona above the disc. 
Overlapped to the continuum, there is also a strong emission line at 6.4 keV, a noticeable narrow feature corresponding to the K$\alpha$ transition of iron,
and a reflection component, usually referred to as \textit{Compton hump}, around $30 \, \rm keV$ \citep{Ghisellini1994, Fiocchi2007}.
The Fe-K$\alpha$ line is attributed to fluorescence in the inner part of the accretion disc, $\sim$ few Schwarzschild radii from the central BH, while the Compton hump is due to Compton-down scattering of high energy photons by high column density reflector $N_{\rm H} \gtrsim 10^{24} \rm \, cm^{-2}$.
Finally, the IB extends from $\sim 1$ $\mu m$ to $\sim 100$ $\mu$m, and it is thought to arise from reprocessed BBB emission by dust. 

In this section, we will focus on the emission in the optical/UV and X-ray bands\footnote{The normalization of the final SED is $L_{\rm bol}$, computed for each active galaxy simulated in \textsc{GAMETE/QSOdust} (see P16 for details).}.


\subsection{Modeling the primary emission}\label{section primary}

\begin{figure}
\centering
\includegraphics[width=8cm]{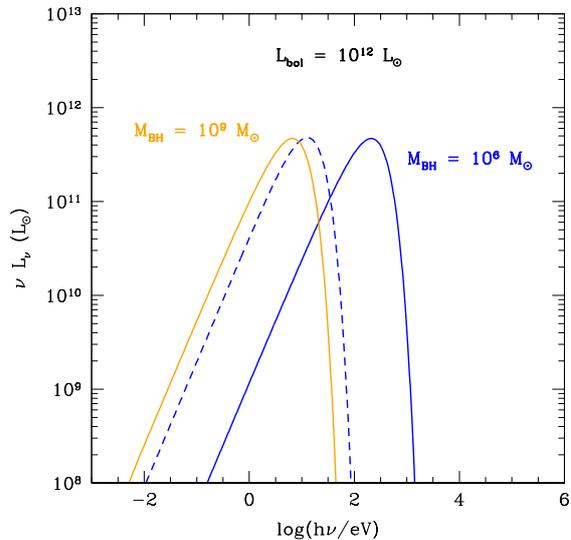}
\caption{ Examples of thermal emission spectra for BHs with masses of $10^6 M_{\odot}$ (blue lines)  and $10^9 M_{\odot}$ (orange line) normalized to a common  bolometric luminosity of $L_{\rm bol} = 10^{12} L_\odot$. Standard thin disc and slim disc models are shown with solid and dashed lines, respectively. For this luminosity, we find that $r_0 > r_{\rm pt}$ for the $10^9 M_{\odot}$ BH so that the slim and the thin disc
models lead to the same emission spectrum. }
\label{figslim}
\end{figure}

We parametrize the emission from the hot corona as a power law
\begin{equation}
L_{\nu} \propto {\nu}^{-\Gamma + 1}e^{h\nu/E_c} ,
\end{equation}
\noindent where $E_c = 300\, \rm keV$ is the exponential cut-off energy \citep{Sozonov2004, Yue2013} and $\Gamma$ is the photon index. We include the reflection component using the PEXRAV model \citep{Magdziarz1995} in the XSPEC package, assuming an isotropic source located above the disc, fixing the reflection solid angle to $2\pi$, and the inclination angle to $60^{\circ}$.
Observations show evidence of a dependence of the photon index $\Gamma$ of the X-ray spectrum on the Eddington ratio $\lambda_{\rm Edd} = L_{\rm bol}/L_{\rm Edd}$ \citep{Grupe2004, Shemmer2008, Zhou2010, Lusso2010, Brightman2013}.
Despite this correlation seems to be found in both the soft and hard bands, the measures of $\Gamma_{\rm 0.5-2keV}$ can be contaminated by the presence of the soft excess, hampering any strong claim of a correlation between the primary emission in this band and $\lambda_{\rm Edd}$. Instead, this contamination is less important in the hard band $[2 - 10]\rm keV$.
\citet{Brightman2013} measured the spectral index $\Gamma_{\rm 2-10 keV}$ of radio-quiet AGNs with $\lambda_{\rm Edd} \lesssim 1$ up to $z \sim 2$, finding that:

\begin{equation}\label{gammabrig}
\Gamma_{\rm 2-10 keV} = (0.32 \pm 0.05) \log\lambda_{\rm Edd} + (2.27 \pm 0.06).
\end{equation}
Here we adopt the above relation to model the dependence of the X-ray spectrum on $\lambda_{\rm Edd}$.

We assume the primary emission in the optical/UV bands to be described as the sum of a multicolour black body spectrum $L^{\rm BB}_{\nu}$, emitted by different parts at different disc temperatures $T$:

\begin{equation}
L^{\rm BB}_{\nu} = L_{0} \int^{T_{\rm max}}_{0} B_{\nu}(T) \left( \frac{T}{T_{\rm max}}\right)^{-11/3}  \frac{dT}{T_{\rm max}},
\end{equation}

\noindent where $B_{\nu}(T)$ is the Planck function and $L_{0}$ is a normalization factor.
The temperature profile of a steady-state, optically thick, geometrically thin accretion disc is \citep{Shakura1973}:

\begin{equation}
\label{lambda}
T(r) = \left( \frac{3GM_{\rm BH} \dot{M}}{8\pi\sigma r^3} \right)^{1/4} \left( 1 - \sqrt{\frac{r_0}{r}}\right)^{1/4},
\end{equation}

\noindent where $M_{\rm BH}$ is the mass of the compact object, $\dot{M}$ the gas accretion rate, $\sigma$ is the Stefan-Boltzman constant and $r_{0}$ is the Innermost Stable Circular Orbit (ISCO), that we assume to be the ISCO for a non-rotating BH. The maximum  temperature $T_{\rm max}$ is achieved at a radius $r(T_{\rm max}) = \frac{49}{36}r_0 $.

Hence, the SED depends both on $\lambda_{\rm Edd}$ and $M_{\rm BH}$. In fact, for a given luminosity,
the peak of the SED is shifted towards higher energies for lower $M_{\rm BH}$ (see Figure \ref{figslim}).
However, the assumption of a standard \textit{thin} disc model is valid when the disc is geometrically thin, i.e. for luminosities below $\sim 30 \%$ of Eddington luminosity. Above this value, the radiation pressure causes an inflation of the disc \citep{McClintock2006}.
Optically thick disc with high accretion rates are better described by \textit{slim} accretion disc models \citep{Abramowicz1988, Sadowski2009, Sadowski2011}, where the photon trapping effect has an important role. In fact, photons produced in the innermost region of the disc are trapped within it, due to large Thompson optical depth, and advected inward. 
The typical radius within which photons are trapped, $r_{\rm pt}$, can be obtained by imposing that the photon diffusion time scale is equal to the accretion time scale, so that \citep{Ohsuga_PT2002}:

\begin{equation}
r_{\rm pt} = \frac{3}{2} R_{s}(\dot{M}/\dot{M}_{\rm Edd,1}) \rm h,
\end{equation}

\noindent where $R_{s} = 2GM_{\rm BH}/c^2$ is the Schwarzschild radius, $\dot{M}_{\rm Edd,1}$ is the Eddington accretion rate and $\rm h=H/r$ is the ratio between the half disc-thickness $\rm H$ and the disc radius $\rm r$. Since $\rm h \approx 1$ in radiation pressure dominated regions, we assume $\rm h = 2/3$ so that $r_{\rm pt} = R_{s}(\dot{M}/\dot{M}_{\rm Edd,1})$. 
Photon trapping causes a cut-off of the emission at higher temperatures and, thus, a shift of the spectrum towards lower energies. 
To consider this feature of super-critical, advection-dominated energy flows, we assume that the radiative emission contributing to the spectrum is that emerging from $r>r_{\rm pt}$. Under this assumption, the difference between \textit{thin} and \textit{slim-like} discs will appear for $L \gtrsim 0.3 L_{\rm Edd}$.

In Figure \ref{figslim} we show the thermal emission corresponding to a bolometric luminosity of $L_{\rm bol} = 10^{12} L_\odot$ and two BH masses $M_{\rm BH} = 10^9 M_{\odot}$ (orange) and $M_{\rm BH} = 10^6 M_{\odot}$ (blue). We compare the classical \textit{thin }disc (solid lines) to that of \textit{slim }disc (dashed line). 
If we consider \textit{thin} discs, for a given $L_{\rm bol}$, BHs with higher masses have a SED which peaks at lower energies.
As a result of photon trapping, a comparable shift towards lower energies is obtained by a $\sim 10^6 \, M_{\odot}$ BH with a super-critical accretion disc, for which $r_{\rm pt} > r_0$.

The relative amplitude of the spectrum in the UV and X-ray bands is usually quantified by the the optical to X-ray spectral index $\alpha_{\rm OX}$, defined as $\alpha_{\rm OX} = -0.384 \log(L_{\rm 2 keV}/L_{2500{\AA}})$. 
Observations \citep{Steffen2006, Just2007, Young2009, Lusso2010, Lusso2016} suggest that $\alpha_{\rm OX}$ increases with $L_{\rm 2500}$, implying that the higher is the emission in the UV/optical band, the weaker is the X-ray component per unit of UV luminosity.
In a recent study, based on a sample of AGNs with multiple X-ray observations at $0 \lesssim z \lesssim 5$, \citet{Lusso2016} found that $\log L_{\rm 2keV} = 0.638 \log L_{2500 {\AA} } + {7.074}$, which implies,

\begin{equation}\label{alpha2016}
\alpha_{\rm OX,2016} = 0.14\log L_{\rm 2500 {\AA}} - 2.72.
\end{equation}

\noindent In what follows, we adopt this relation to quantify the relative contribution of the optical/UV and X-ray spectrum, and truncate the emission from the hot corona at energies below $\sim 3 T_{\rm max}$.

\subsection{Absorbed spectrum}\label{abs}

\begin{figure}
\centering
\includegraphics[width=8cm]{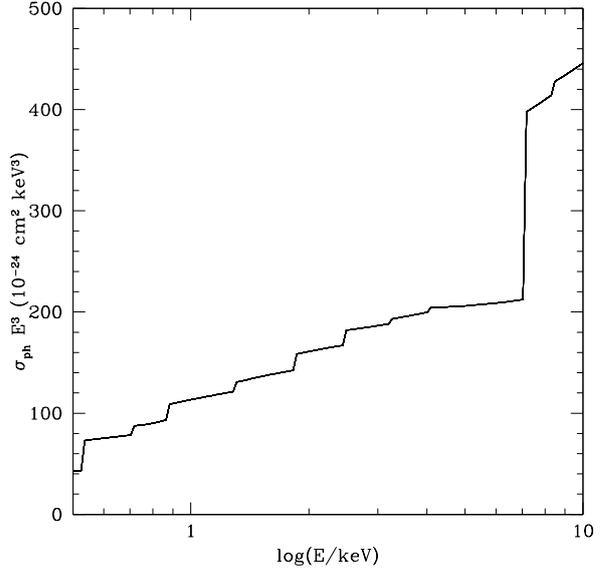}
\caption{Photoelectric cross section as a function of energy for $Z = Z_\odot$.}
\label{cross}
\end{figure}

The radiation produced from the accreting process can interact with the gas and dust in the immediate surroundings of the BH. For the purpose of this study, we consider only the absorption in the X-ray band. 
The two main attenuation processes are photoelectric absorption and Compton scattering of photons against free electrons.
The effect of these physical processes is to attenuate the intrinsic flux, $F_{\nu}$, by:

\begin{equation}
F^{\rm obs}_{\nu} = F_{\nu}e^{-\tau_\nu}.
\end{equation}

\noindent At $h\nu \gtrsim 0.1$ keV and under the assumption of a fully-ionized H-He mixture, the optical depth $\tau_\nu$ can be written as $\tau_\nu = (1.2\sigma_T + \sigma_{ph})N_{H}$ \citep{Yaqoob1997} where $N_{H}$ is the hydrogen column density and $\sigma_T$ and $\sigma_{ph}$ are the Thomson and the photoelectric cross section, respectively.

\citet{Morrison1983} computed an interstellar photoelectric absorption cross section $\sigma^{Z_{\odot} }_{ph}$ as a function of  energy in the range [0.03-10]~keV, for solar metallicity $Z_{\odot}$\footnote{We have renormalized $\sigma_{\rm ph}$ that \citealt{Morrison1983} originally computed for $Z = 0.0263$ to a solar metallicity value of $Z_\odot = 0.013$ \citep{Asplund2009}.}.

In our simulations, the gas metallicities of high-z BH host galaxies span a wide range of values, with $0 \lesssim Z \lesssim Z_{\odot}$. To account of the metallicity dependence of the absorbing material, we separate the photoelectric cross section into its components

\begin{equation}\label{sigma}
\sigma_{ph} = \sigma_{H} + \sigma_{He} + \sigma_{met},
\end{equation}

\noindent where $\sigma_{H}$ and $\sigma_{He}$ represent the contribution of hydrogen and helium.

The hydrogen ionization energy $\sim 13.6 \rm eV$ and helium second ionization energy $\sim 54.4 \rm eV$ are much lower than the energy in the X-ray band ($\sim \rm keV$), hence $\sigma_{H}$ and $\sigma_{He}$ can be safely evaluated in Born approximation.
Following \citet{Shu1991}, the cross section in Born approximation for a hydrogen atom is

\begin{equation}
\sigma_{\rm X} = \frac{8 \pi}{3 \sqrt{3}} \frac{Z^4_{\rm X} m_e e^{10}}{c \hbar^3(\hbar \omega)}\sqrt{ \frac{48 Z_{\rm X} e^2}{2 a_Z \hbar \omega}} ,
\end{equation}

\noindent where $Z_{\rm X}$ is the atomic number for the X-th element (1 for H, 2 for He), $m_e$ and $e$ are the electron mass and charge, $c$ is the speed of light, $\hbar$ the reduced Plank constant and $a_Z = \hbar/Z_{\rm X}m_e e^2$.

\begin{figure*}
\includegraphics[width=5.6cm]{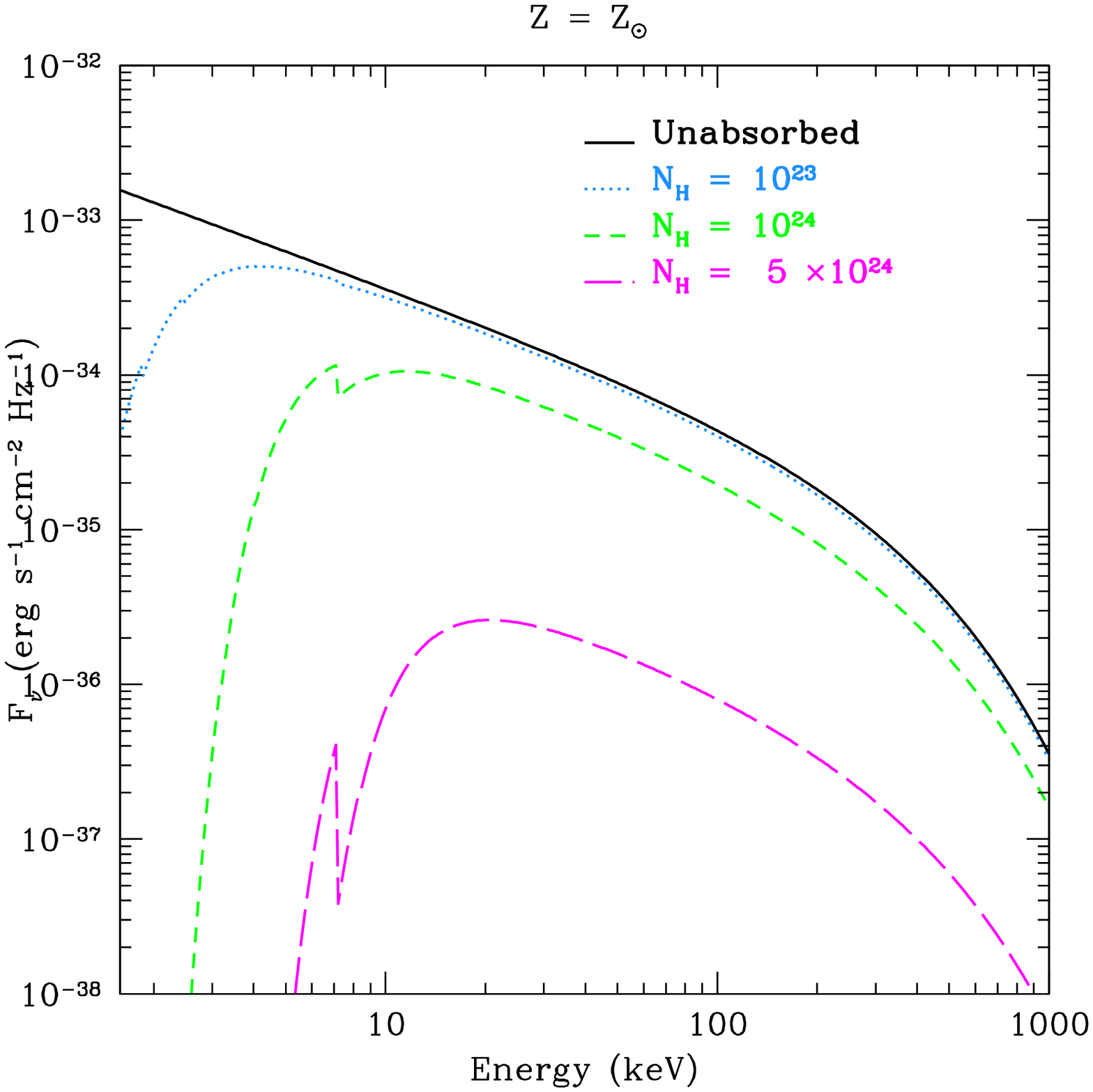}
\includegraphics[width=5.6cm]{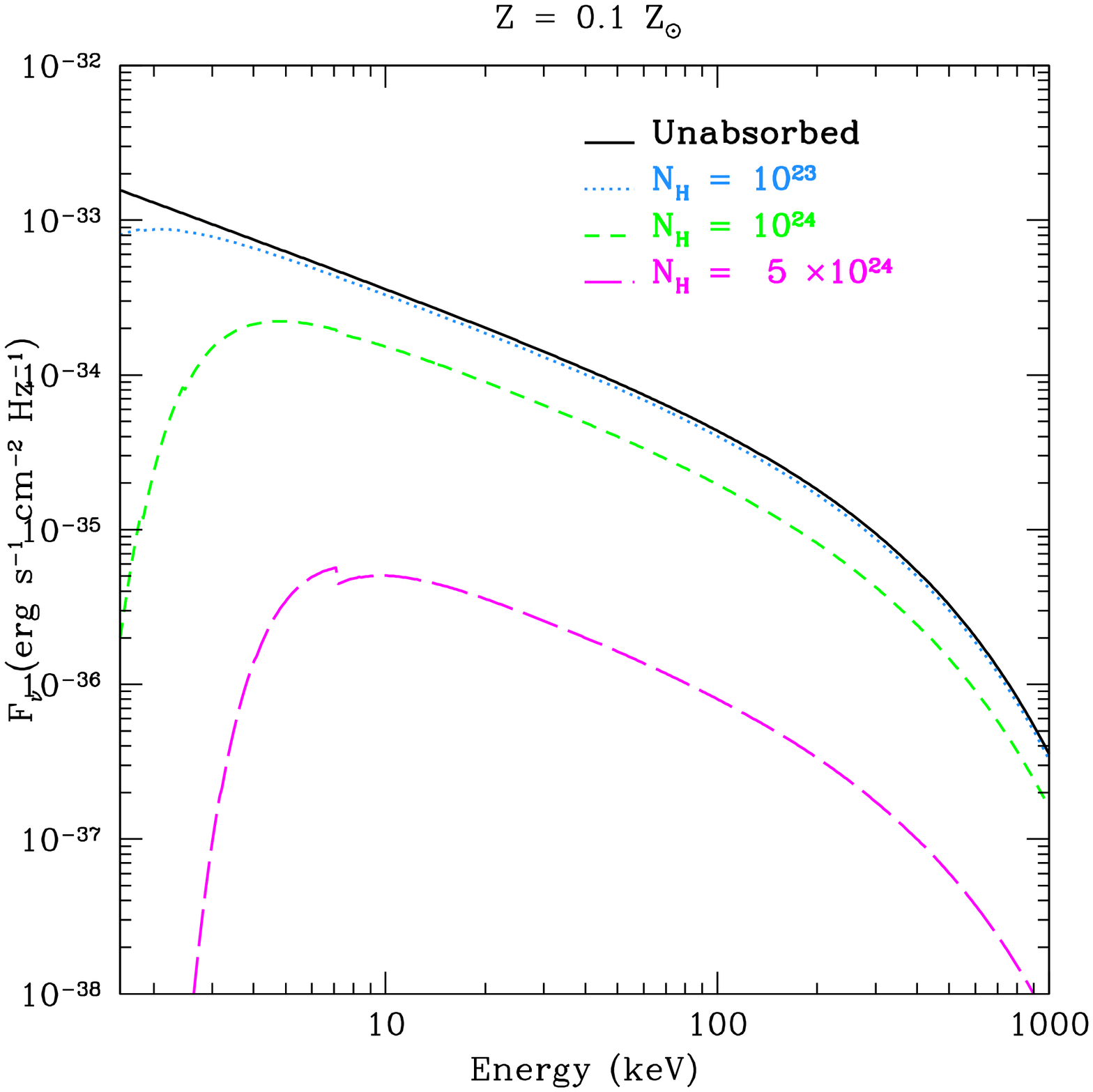}
\includegraphics[width=5.6cm]{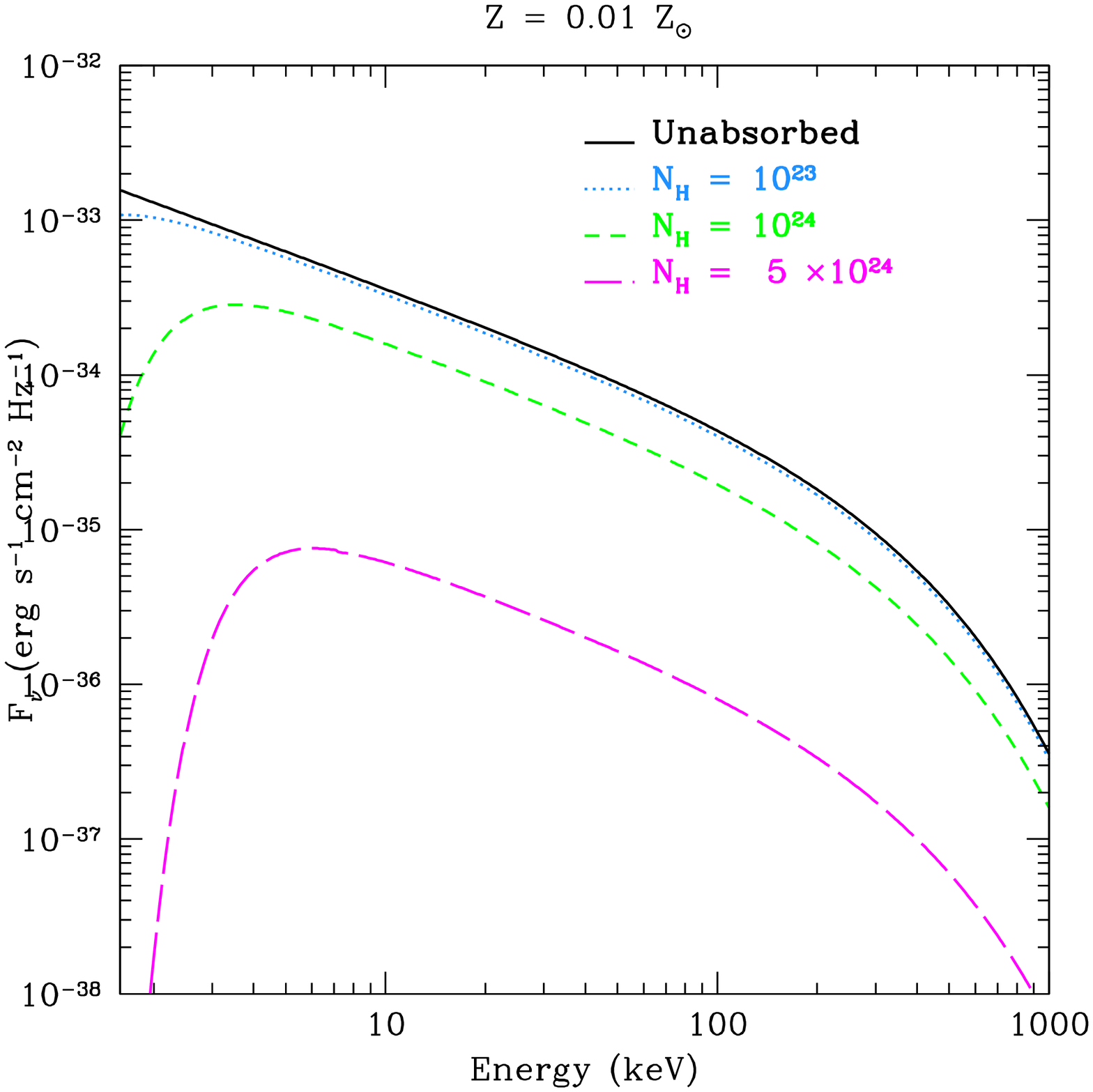}
\caption{Primary (black solid line) and reprocessed emissions (dashed lines) of accreting BHs for column densities $N_{\rm H} = (10^{23}$, $10^{24}$, $5 \times 10^{24}$) $\rm cm^{-2}$. Different panels refer to different metallicities: $Z = Z_\odot$ (left), $Z = 0.1Z_\odot$ (middle) and $Z = 0.01Z_\odot$ (right).}
\label{abso}
\end{figure*}

In Figure \ref{cross} we can see the photoelectric cross section for $Z = Z_{\sun}$. 
For energies $\gtrsim 0.2 \rm keV$, $\sigma_{\rm ph}$ is dominated by metals, in particular C and N. 
The cross section presents several gaps that correspond to the K-shell energies of different elements.
In fact, in the evaluation of $\sigma_{\rm ph}$  it has been taken into account that an element X contributes to the absorption only if the photon energy is greater than the K-shell energy, with the highest energy gap corresponding to Fe.
The photoelectric cross section decreases for increasing energy, when the Thomson cross section $\sigma_{\rm T}$ becomes dominant (for $E \gtrsim 10$ keV at $Z = Z_\odot$).
Thus, softer X-ray photons are expected to be more absorbed than harder ones.
This feature is well visible in Figure \ref{abso}, where the intrinsic spectrum for $L_{\rm bol} = 10^{12} L_{\odot}$ and $M_{\rm BH} = 10^9 M_{\odot}$ (black line) is compared to the spectra attenuated by gas with $Z = Z_\odot$, $0.1 \, Z_\odot$ and $0.01 \, Z_\odot$ (from left to right respectively) and different values of hydrogen column density $N_{\rm H}$ (dashed lines), that have been computed consistently with the diffuse and cold gas density profiles (see Section \ref{sample}). The effect of metallicity is relevant only at lower energies, where the photoelectric cross section is dominant. As already discussed, in fact, at energies $E \gtrsim 10$ keV the Thomson cross section becomes dominant, removing the absorption dependence on metallicity.\\
Compton thick AGNs, which are usually characterized by $N_{\rm H} \gtrsim 1.5 \times 10^{24} \, \rm cm^{-2}$, are completely absorbed in the soft band. The emission peak moves to $\sim 20\rm\, keV$, and the corresponding magnitudes is $\sim$ 2 orders of magnitude lower than in the intrinsic spectrum.
For $N_{\rm H} \lesssim 10^{25} \, \rm cm^{-2}$, the direct emission is visible at energies $E \gtrsim 10 \rm \, keV$, and they are labelled as \textit{transmission-dominated} AGNs. For even larger column densities ($N_{\rm H} > 10^{25} \, \rm cm^{-2}$) direct X-ray emission is strongly affected by Compton scattering and fully obscured, and only the faint reflection component can be detected (\textit{reflection-dominated} AGNs).
We note, however, that X-ray observations of $z \gtrsim 4$ quasars typically sample the rest-frame hard X-ray band.\\
The condensation of the absorbing material into grains reduces the value of $\sigma_{\rm ph}$. \citet{Morrison1983} estimate the importance of this effect, evaluating the photoelectric cross section in the case that all the elements but H, He, Ne and Ar are depleted in grains, with the exception of O, for which the condensation efficiency is assumed to be 0.25.
The variation in the photoelectric cross section is relatively modest, $\sim 11\%$ at $E \sim 0.3$ keV and $\sim 4$\% at 1 keV.   Hence, hereafter we neglect this effect.\\
Despite we are restricting our analysis to the X-ray part of the emission spectrum, it is important to note that the 
absorbed radiation will be re-emitted at lower energies. \citet{Yue2013} find that for Compton-thick systems, secondary photons emitted by free-free, free-bound and two-photon processes can 
increase the luminosity by a factor of $\sim 10$ in the rest-frame $[3 - 10]$~eV, which will be redshifted in the near IR at $z=0$. 
As a result, most of the energy emitted is expected to be observed in the IR and soft-X-ray bands \citep{Pacucci2015, Pacucci2016, Natarajan2016}. 

\section{The sample}\label{sample} 

In Section \ref{Sec1} we have introduced our emission model for accreting BHs. 
Physical inputs required to compute the spectrum are the BH mass, $M_{\rm BH}$, the bolometric luminosity, $\rm L_{\rm bol}$, the Eddington accretion ratio, $\dot{M}/\dot{M}_{\rm Edd,1}$, the metallicity, $Z$, and the column density, $N_{\rm H}$. We adopt the semi-analytic model \textsc{GAMETE/QSOdust}, in the version described by P16, to simulate these properties for a sample of BH progenitors of $z \gtrsim 6$ SMBHs. In this section, we
first summarize the main properties of the model and then we describe the physical properties of the simulated sample.

\subsection{Simulating SMBH progenitors with \textsc{GAMETE/QSOdust}}

\begin{figure}
\centering
\includegraphics[width=8cm]{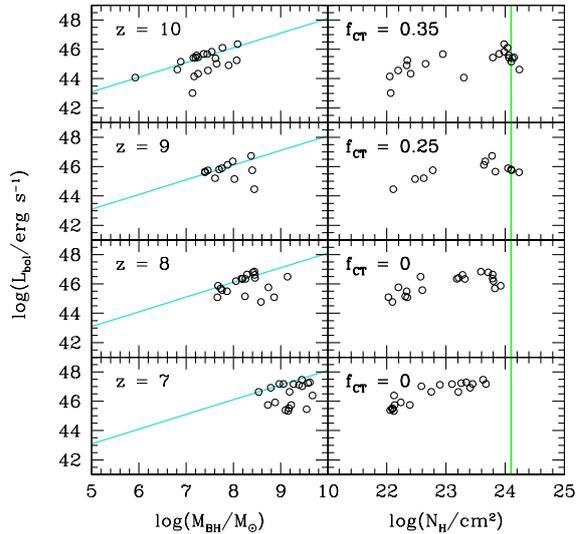}
\caption{Properties of BH progenitors extracted from 30 simulations at $z = 7, 8, 9$ and 10. Bolometric luminosities are shown as a function of BH masses (\textit{left panel}) and hydrogen column density in the host galaxy $N_{\rm H}$ (\textit{right panels}). Cyan lines represent $L_{\rm Edd}(M_{\rm BH})$. The green vertical line represents the $N_{\rm H}$ corresponding to a Compton-thick system, while $f_{\rm CT}$ is the fraction of Compton-thick BHs present at that redshift.}  
\label{propertyBH}
\end{figure}

The code allows to reconstruct several independent merger histories of a $10^{13} M_\odot$ DM halo assumed to host a typical $z \sim 6$ SMBH, like J1148 (e.g. \citealt{Fan2004}).
The time evolution of the mass of gas, stars, metals and dust in a two-phase interstellar medium (ISM) is self-consistently followed inside each progenitor galaxy. 
The hot diffuse gas, that we assume to fill each newly virialized DM halo, can gradually cool through processes that strongly depend on the temperature and chemical composition of the gas. For DM halos with virial temperature $T_{\rm vir} < 10^4$~K, defined as \textit{minihalos}, we consider the contribution of $\rm H_2$, OI and CII cooling \citep{Valiante2016}, while for  Ly$\alpha$-halos ($T_{\rm vir} \geq 10^4$~K) the main cooling path is represented by atomic transitions. In quiescent evolution, the gas settles on a rotationally-supported disc, that can be disrupted when a major merger occurs, forming a bulge structure. 
The hydrogen column density $N_{\rm H}$ has been computed taking into account the gas distribution in the diffuse and cold phases. 
We assumed a spherically-symmetric Hernquist density profile for the gaseous bulge \citep{Hernquist1990},
\begin{equation}
\rho_b(r) = \frac{M_b}{2 \pi}\frac{r_b}{r(r+r_b)^3},
\end{equation} 
\noindent where $M_b$ is the bulge mass of the gas, $r_b$ is the scale radius $r_b = R_{\rm eff}/1.8153$ \citep{Hernquist1990}, 
and the effective radius, $R_{\rm eff}$, has been computed as $\log(R_{\rm eff}/{\rm kpc}) = 0.56\log(M_b + M_b^\star) - 5.54$, where $M_b^\star$ is the stellar mass of the bulge  \citep{Shen2003}.
For the diffuse gas, we adopt an isothermal density profile (see Section 2.1 and 2.2 in P16) and we do not consider the contribution of the galaxy disc to the absorbing column density.\\

We assume BH seeds to form with a constant mass of $100 \, M_\odot$ as remnants of Pop~III stars in halos with $Z \leq Z_{\rm cr} = 10^{-4} \, Z_\odot$. 
As a result of metal enrichment, BH seeds are planted in halos with a mass distribution peaking around $M_{\rm h} \sim 10^7 \, M_{\odot}$, at $z > 20$, below which no Pop~III stars is formed. 

The BH grows via gas accretion from the surrounding medium and through mergers with other BHs. 
Our prescription allows to consider quiescent and enhanced accretion following merger-driven infall of cold gas, which loses angular momentum due to torque interactions between galaxies.
We model the accretion rate to be proportional to the cold gas mass in the bulge $M_{\rm b}$, and inversely proportional to the bulge dynamical time-scale $\tau_{\rm b}$:

\begin{equation}
\dot{M}_{\rm accr} = \frac{f_{\rm accr} M_{\rm b}}{\tau_{\rm b}},
\end{equation}

\noindent where $f_{\rm accr} = \beta f(\mu)$, with $\beta = 0.03$ in the reference model and $f(\mu) = \max[1, 1+2.5(\mu - 0.1)]$, so that mergers with $\mu \leq 0.1$ do not trigger bursts of accretion. 

As discussed in Section \ref{section primary}, once the accretion rates become high, the standard \textit{thin} disc model is no longer valid.
Therefore, the bolometric luminosity $L_{\rm bol}$ produced by the accretion process has been computed starting from the numerical solution of the relativistic slim accretion disc obtained by \citet{Sadowski2009}, adopting the fit presented in \citet{Madau2014}. This model predicts mildly super-Eddington luminosities even when the accretion rate is highly super-critical.

The energy released by the AGN can couple with the interstellar gas. We consider energy-driven feedback, which drives powerful galactic-scale outflows, and SN-driven winds, computing the SN rate explosion for each galaxy according to formation rate, age and initial mass function of its stellar population \citep{deBennassuti2014,Valiante2014}. 

Finally, in BH merging events, the newly formed BH can receive a large center-of-mass recoil due to the net linear momentum carried by the asymmetric gravitational wave \citep{Campanelli2007,Baker2008} and we compute the \textit{kick} velocities following \citet{Tanaka2009}.

We refer the reader to P16 for a more detailed description of the model.

\subsection{Physical properties of the sample}

\begin{figure}
\centering
\includegraphics[width=8cm]{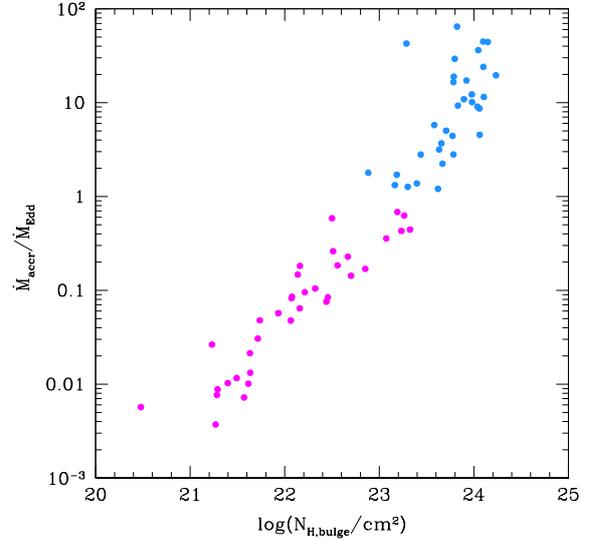}
\caption{Column density of the bulge and Eddington accretion ratio for each of the active BHs found at $z = 7, 8, 9, 10$. Azure (magenta) represents super- (sub-) critical accreting BHs, i.e. those for which $\dot{M}/\dot{M}_{\rm Edd} > 1$}
\label{propertymedd}
\end{figure}

We run $N_r$ independent merger trees and reproduce all the observed properties of one of the best studied quasars, SDSS J1148+5152 (hereafter J1148) at $z=6.4$ that we consider as a prototype of luminous $z \gtrsim 6$ quasars.
We choose $N_r = 30$ to match the statistics of the currently known sample of $z \gtrsim 6$ quasars with robust BH mass measurements and $M_{\rm BH} \gtrsim 10^9 M_\odot$ \citep{Fan2001,Fan2003,Fan2004,Fan2006}.

Figure \ref{propertyBH} shows the bolometric luminosity as a function of the BH mass (left panel) and hydrogen column density (right panel) for \textit{active} BH progenitors (i.e. with $\lambda_{\rm Edd} \geq 5 \times 10^{-3}$) of SMBHs extracted from the simulations at $z = 7,8,9,10$. All BH progenitors have masses $M_{\rm BH} \gtrsim 10^{6} M_{\odot}$ and bolometric luminosities $L_{\rm bol} \gtrsim 10^{42}$ erg/s. 
As it can be seen from the figure, luminosities never exceed $\sim$ few $L_{\rm Edd}$ (cyan lines), also for super-critical accreting BHs. This is a result of the low radiative efficiencies of the \textit{slim} disc solution: only a small fraction of the viscosity-generated heat can propagate, while the larger fraction is advected inward. In the right panel of the figure, we show the relation between hydrogen column density $N_{\rm H}$ and bolometric luminosity. At all redshifts, our sample is composed only by \textit{transmission-dominated} AGNs.
The vertical lines indicate the column density above which the systems are classified as Compton-thick. 
The fraction of Compton-thick AGNs, $f_{\rm CT}$, is also shown. We find that $f_{\rm CT}$ increases with redshift, 
ranging between $35\%$ at $z = 10$  to $\sim 0 $ at $z = 7$ and that $f_{\rm CT} \sim 45 \%$ for all
the simulated sample at all redshifts. These numbers are consistent with the loose limits inferred from the analysis of the cosmic
X-ray background (CXB) with AGN population synthesis models, which generally find $f_{\rm CT} = 5 - 50 \%$ 
\citep{Ueda2003, Gilli2007, Akylas2012}, and with indications of growing obscuration with redshift \citep{LaFranca2005, Treister2009,
Brightman2012} and luminosity (\citealt{Vito2013}, see however \citealt{Buchner2015}).

The environmental conditions in which these BHs grow play an important role in determining the accretion regime. Figure \ref{propertymedd} shows the Eddington accretion ratio $\dot{M}/\dot{M}_{\rm Edd}$, where $\dot{M}_{\rm Edd} = 16L_{\rm Edd}/c^{2}$, as a function of the hydrogen column density of the bulge, which provides the gas reservoir to BH accretion. 
We find a positive correlation of the ratio with $N_{\rm H,bulge}$, showing that, when $N_{\rm H,bulge} \gtrsim 10^{23} \rm cm^2$, BHs accrete at super-critical rates.

\begin{figure}
\centering
\includegraphics[width=8cm]{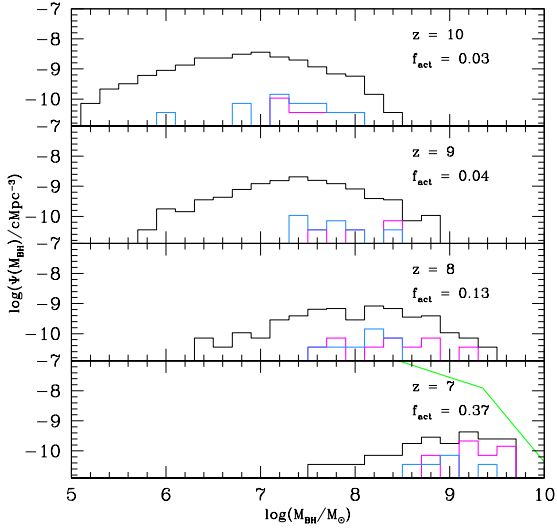}
\caption{The mass function of BH progenitors at four different snapshots (z = 10, 9, 8 and 7 from top to bottom). The black line shows the total while the azure solid and magenta dotted lines indicate active BHs accreting at super and sub-Eddington rates, respectively. The fraction of active BHs at each redshift, $f_{\rm act}$, is also reported. {The green solid line in the bottom panel represents the BH mass function inferred from observations by \citet{Willott2010} at $z = 6$}.}
\label{density}
\end{figure} 

\begin{figure}
\centering
\includegraphics[width=8.3cm]{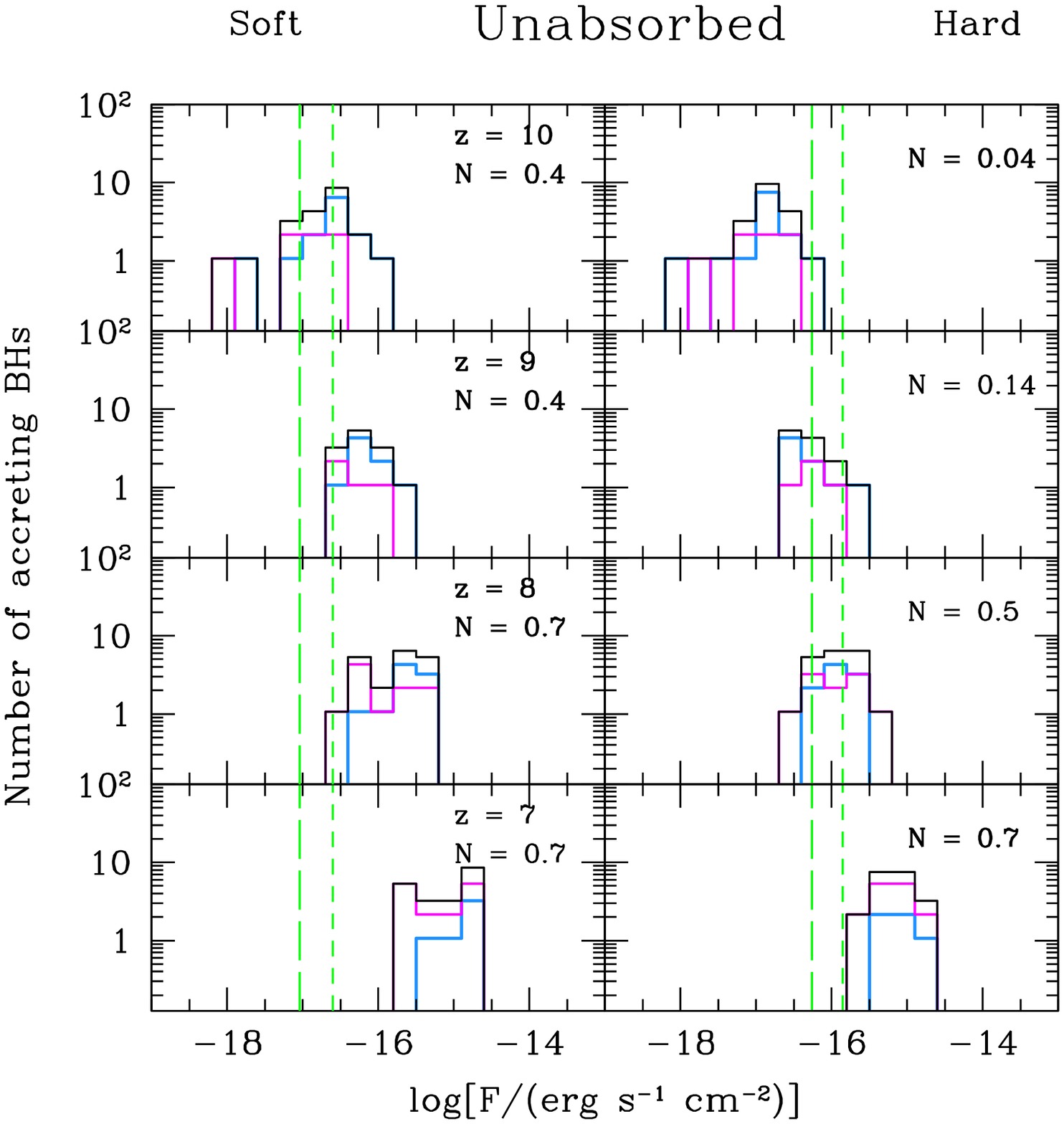}
\includegraphics[width=8.3cm]{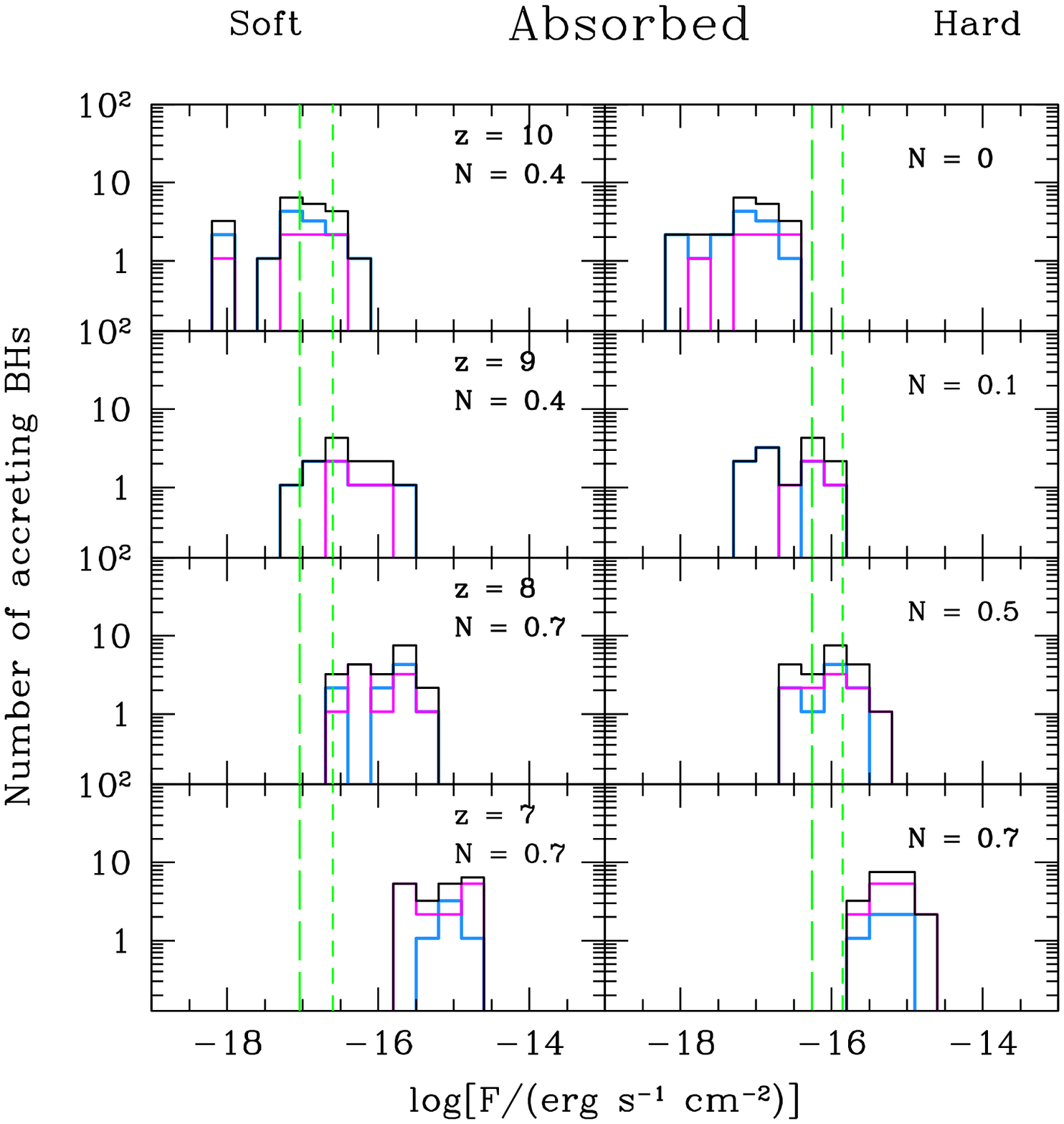}
\caption{Flux distribution for each snapshot (black solid lines), divided in super- (azure) and sub- (magenta) Eddington accreting BH progenitors. We report both the \textit{unabsorbed} model (\textit{top panel}) and the \textit{absorbed} model (\textit{bottom panel}), for the soft (left panels) and hard (right panels) \textit{Chandra} bands. Vertical dashed green lines represent different \textit{Chandra} flux limits: CDF-S 4 Ms (long-dashed, \citealt{Xue2011}), $F_{\rm CDF-S} = 9.1 \times 10^{-18}$ ($5.5 \times 10^{-17}$) $\rm erg \, s^{-1}\,cm^{-2}$ and CDF-N 2 Ms (short-dashed, \citealt{Alexander2003}), $F_{\rm CDF-N} = 2.5 \times 10^{-17}$ ($1.4 \times 10^{-16}$) $\rm erg \, s^{-1}\,cm^{-2}$ in the soft (hard) band. In each panel, we also show the average number N of active progenitors with flux larger than CDF 4 Ms flux limit.}
\label{fluxes}
\end{figure}

In the current model we do not take into account possible anisotropy of the AGN structure, such as the presence of a cleaned (dust and gas free) region from which the nucleus can be visible. For this reason we will investigate two extreme scenarios: the first assumes that there is no important absorption and that the observed X-ray emission is the intrinsic one (\textit{unabsorbed} case), while in the second we compute the absorption as explained in Section \ref{abs} (\textit{absorbed} case).
 
The first important quantity that we can compute is the BH mass function $\Psi(M_{\rm BH})$ of BH progenitors of $z \sim 6$, luminous quasars. Figure \ref{density} shows $\Psi(M_{\rm BH})$ (black line) at different redshifts. The contribution of super- (azure solid) and sub- (magenta dotted) Eddington accreting BHs is also shown. Here the lines represent the averages over 30 merger tree simulations and the comoving volume $V$ of the Universe in which BHs are distributed is $1 \, \rm Gpc^{3}$, as the observed comoving number density of quasars at $z \sim 6$ is $n = 1 \rm \, Gpc^{-3}$ \citep{Fan2004}.
In the the bottom panel of Figure \ref{density}, we compare our results with the BH mass function inferred from observations of SMBHs by \citet{Willott2010} at $z=6$ (shown with the green solid line). As expected, our predictions are below the observed distribution. In fact, our calculations describe the mass functions of BH progenitors of $z = 6$ SMBHs, namely a sub-population of existing BHs. 
This comparison is meant to show that our model predictions do not exceed the observed BH mass function.

At each redshift we consider the whole population of BH progenitors (active and inactive) along the simulated hierarchical merger histories (black solid histogram), with the exclusion of possible satellite BHs and kicked out BHs. These are assumed to never settle (or return) to the galaxy center, remaining always inactive (i.e. they do not accrete gas) and do not contribute to the assembly of the final SMBH (see P16 for details).
The black solid histogram shows that the majority of BHs are temporarily non accreting BHs, due to the reduced gas content in the bulge.
The fraction of active BHs in also reported in Figure \ref{density} for the 4 snapshots. It increases by a factor $\sim 1.3$ from $z=10$ to $z=9$,  
$\sim 3.2$ from $z = 9$ to $z = 8$ and  $\sim 2.8$ from $z = 8$ to $z = 7$. This is due to the increasing fraction of BHs that accrete 
at sub-Eddington rates (see also Fig.~4 in P16).

While the progenitors mass function is relatively flat at $z=7$, a pronounced peak in the distribution becomes visible at higher redshifts, around $M_{\rm BH, peak}\sim 10^7 \, (2.5\times 10^6)$ M$_\odot$ at $z=8 \, (10)$. The mass density, particularly at the low mass end, is shifted towards more massive BHs at $z\leq 8$, as a consequence of BH growth due to mergers and gas accretion. Our simulations are constrained to reproduce the final BH mass of J1148 at $z_{0} = 6.4$, thus the total number of progenitors naturally decreases as an effect of merging (major and minor) and gravitational recoil processes, implying a lower/poorer statistics as redshift approaches $\sim z_0$. 
Finally, the decreasing trend in the number density of $M_{\rm BH}<M_{\rm BH, peak}$ BHs, reflects the effects of chemical feedback. Efficient metal enrichment at $Z\geq Z_{\rm cr} = 10^{-4}\, Z_{\odot}$ inhibits the formation of Pop~III stars and BHs already at $z<20$. At lower redshifts the effects of dust and metal line cooling allows the gas to fragment more efficiently, inducing the formation of lower mass (Pop~II) stars \citep{Schneider2002, Schneider2003, Schneider2012a}. 
As BH seeds grow in mass, the number density at the low-mass end decreases with time.
By $z\sim 7$ the population of $<10^{6}$ M$_\odot$ active progenitors is fully-evolved into more massive objects.
The number and redshift distribution of accreting BHs in the two different accretion regimes have been widely investigated and discussed in P16. The resulting active BH mass functions reflect these properties. Super-Eddington accreting BHs are the dominant component ($> 60\%$) down to $z\sim 10$ as indicated by the azure histogram in the upper panel of Figure \ref{density}. At lower $z$, super-critical accretion becomes progressively less frequent ($<24\%$), and sub-Eddington accretion dominates BH growth down to $z\sim 6-7$.\\

\section{Results and discussion}\label{results}

In this section we analyse the X-ray luminosity of the BH sample introduced in the previous section and we discuss the best observational strategies to detect them by critically assessing the main reasons which have, so far, limited their observability. 

\begin{figure*}
\centering
\includegraphics[width=8cm]{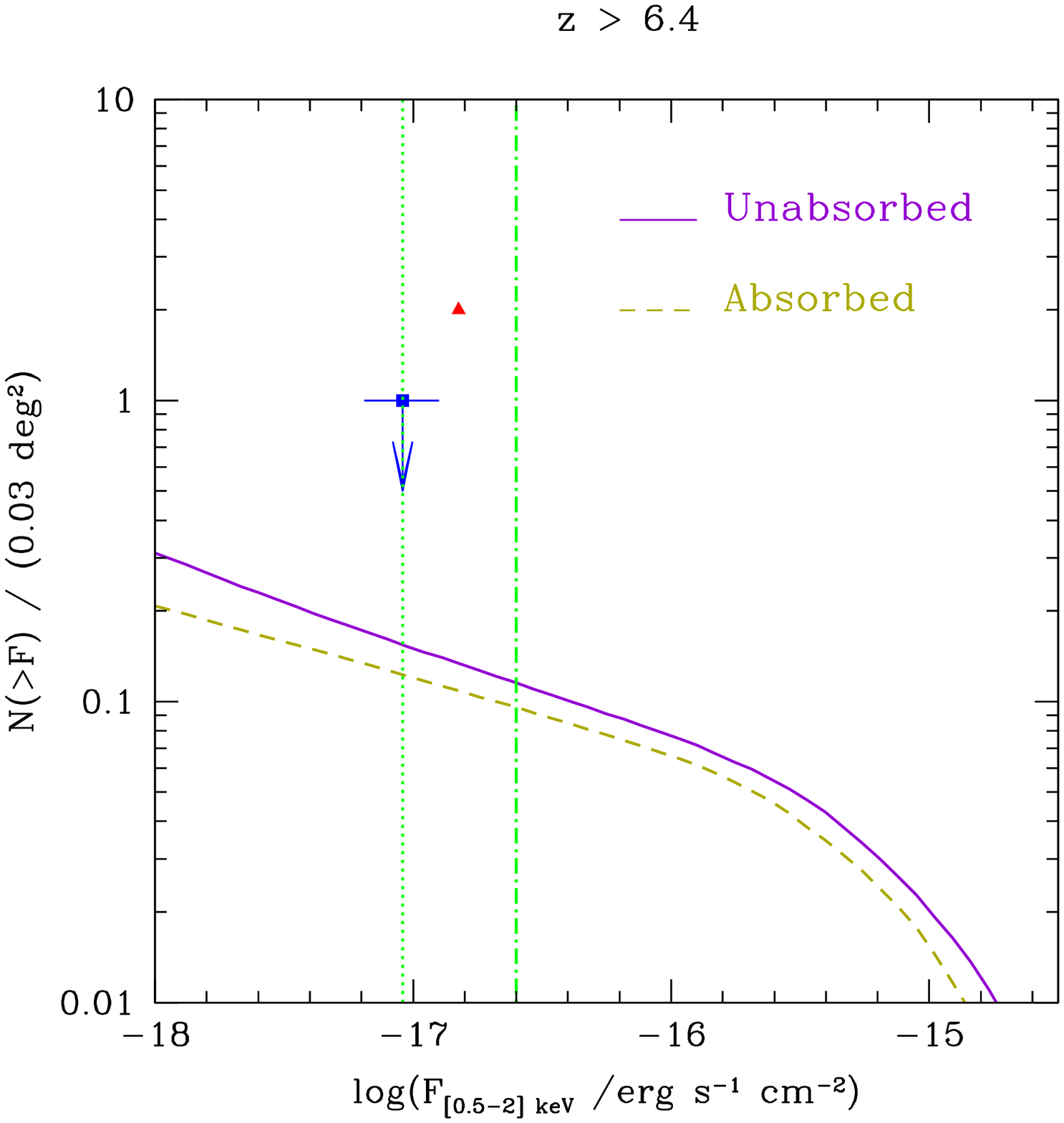}
\includegraphics[width=8cm]{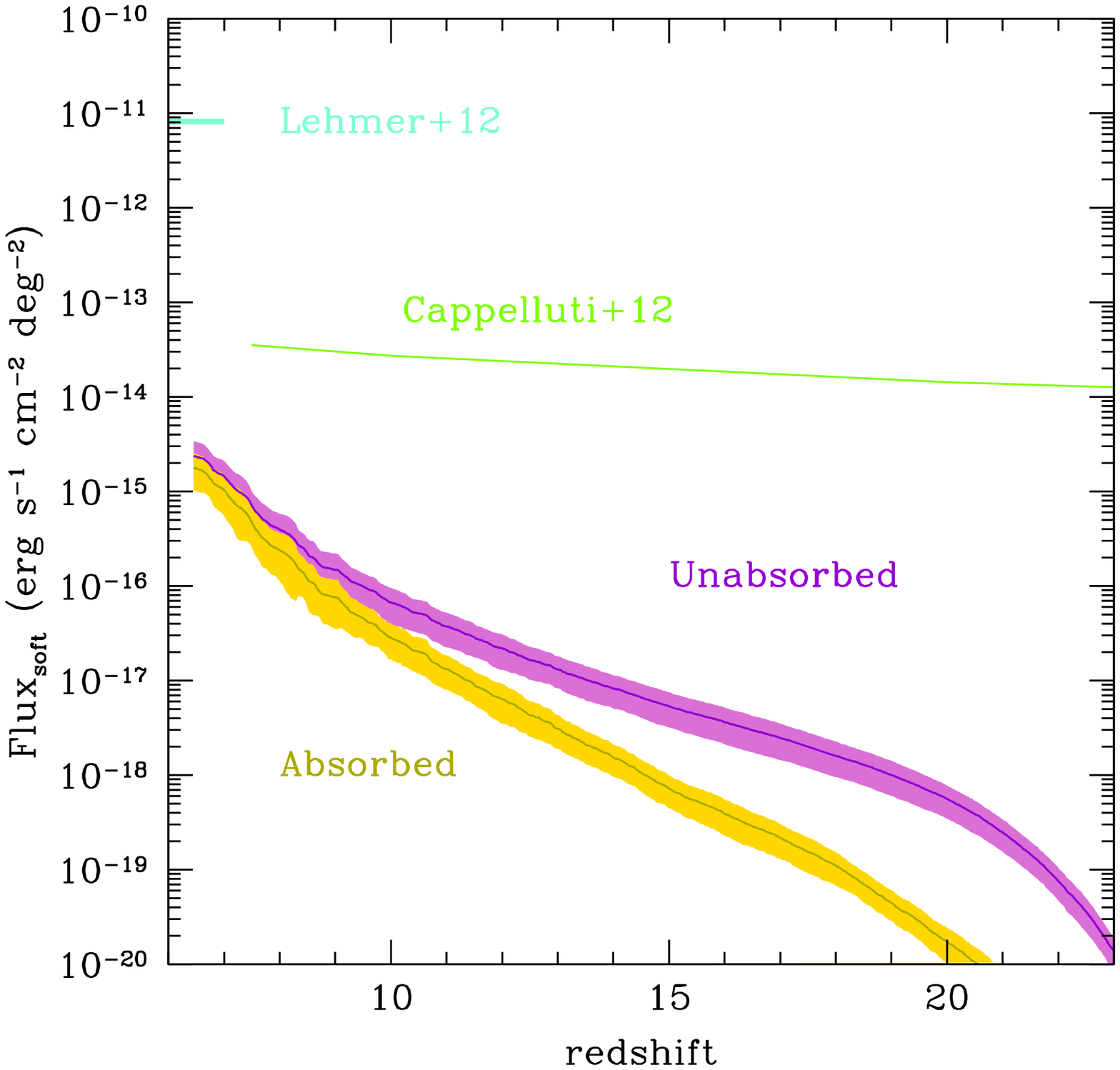}
\caption{{\it Left panel:}Number of active BH progenitors, per unit area of $0.03 \, \rm deg^2$, with a flux larger than F in the \textit{Chandra} soft band, as a function of F. Predictions for the \textit{unabsorbed} (solid violet) and \textit{absorbed} (dashed ochre) models are shown. Vertical green lines represent two different \textit{Chandra} flux limits: CDF-S 4 Ms (dotted lines) and CDF-N 2 Ms (dashed-dotted lines). Red triangle and blue square represent, respectively, the observations obtained by \citet{Giallongo2015} and the upper limit of \citet{Weigel2015}. {\it Right panel:} Cosmic X-ray Background in the soft band [0.5 - 2]~keV predicted by the absorbed and unabsorbed models. The solid lines show the
average among 30 independent simulations and the shaded region is the 1-$\sigma$ scatter. We also show the soft CXB measured by
\citet{lehmer2012} in the 4Ms CDF-S and the upper limit on $z > 7.5$ accreting BHs placed by \citet[see text]{cappelluti2012}.} 
\label{numberflux}
\end{figure*}

\paragraph*{Black hole occupation fraction.}

\noindent The black hole occupation fraction $f_{\rm BH}$ represents the number fraction of galaxies seeded with a BH, regardless the nuclear BHs are active or not. This quantity, not to be confused with the \textit{AGN} fraction, is directly related to the seeding efficiency. 
In this work, we assume that a BH seed is planted once a burst of Pop~III stars occurs in a metal poor, newly virialized halo, as explained in Section \ref{sample}. 
As already mentioned above, in the model we account for the possibility that a galaxy may lose its central BH during a major merger with another galaxy, due to large center-of-mass recoil velocity resulting from net-momentum carrying gravitational wave emission produced by the merging BH pair. 
As a result of this effect, the occupation fraction depends not only on the seeding efficiency, but also on the merger histories of SMBHs.

\citet*{Alexander2014} developed a model in which super-exponential accretion in dense star clusters is able to build a $\sim 10^4 \, M_{\odot}$ BH in $\sim 10^7$ yr, starting from light seeds. 
The subsequent growth of this BH, up to $\sim 10^9 \, M_{\odot}$, is driven by Eddington-limited accretion.
They show that with this mechanism even a low occupation fraction of $f_{\rm BH} \sim 1-5\%$ can be enough to reproduce the observed distribution of $z > 6$ luminous quasars.

However, despite the local BH occupation fraction approaches unity, there are no strong constraints on the value of $f_{\rm BH}$ at high-$z$. In fact, the observed SMBHs number density at $z = 0$ could be reproduced even if $f_{\rm BH} \sim 0.1$ at $z \sim 5$, as a result of multiple mergers experienced by DM halos in the hierarchical formation history of local structures \citep{Menou2001}. 

By averaging over 30 different merger trees, we predict that $f_{\rm BH}$ increases with time, finding an occupation fraction of $f_{\rm BH} = 0.95,\, 0.84, \, 0.76,\,0.70$, at $z = 7, 8, 9, 10$, respectively\footnote{Considering all the simulated galaxies in our sample, at all redshift, we find an occupation fraction of $f_{\rm BH} = 0.35$.}.
Hence, more than $70 \%$ of the final SMBH progenitors host a BH in their centre at $z<10$. Indeed, our simulated $f_{\rm BH}$ is higher than those predicted for average volumes of the Universe, as mentioned above, suggesting that the low occupation fraction is not the main limiting process for the X-ray detectability of BHs at $z > 6$.

\begin{figure}
\centering
\includegraphics[width=9cm]{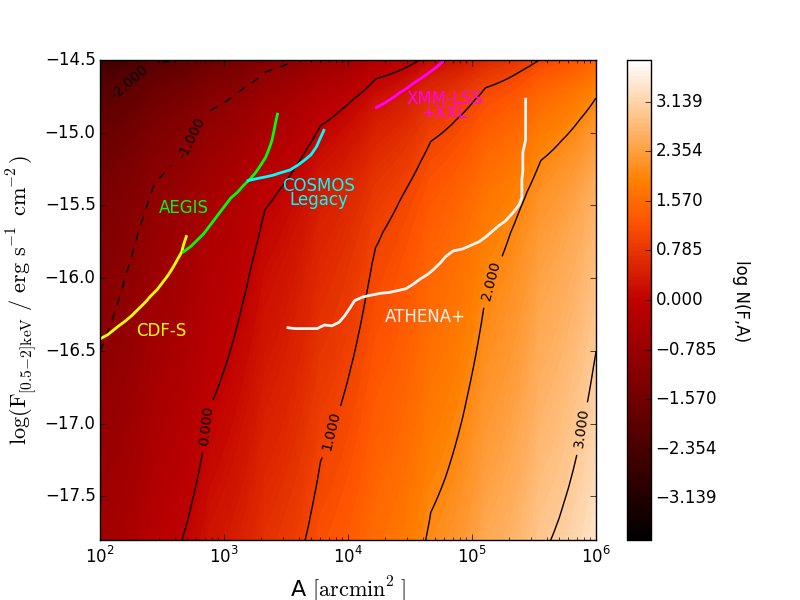}
\includegraphics[width=9cm]{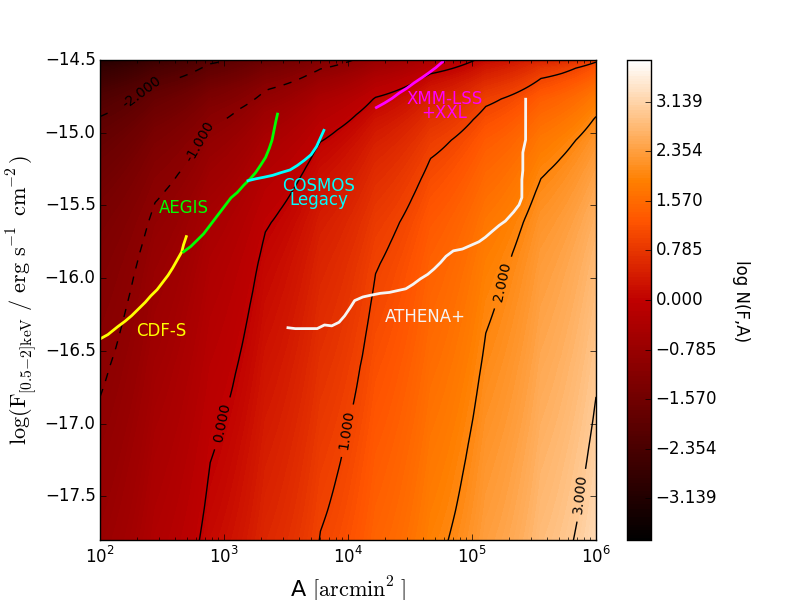}
\caption{Number of progenitors potentially observable in a survey with sensitivity $F_{\rm [0.5-2] keV}$ and probing an area A for the \textit{unabsorbed} (top panel) and \textit{absorbed} (bottom panel) models. Black lines represent the values of $\log N(F, A) = -2, -1$ (dashed lines) and 
$\log N(F, A) = 0, 1, 2$ and 3 (solid lines). We also show the area/flux coverage achieved by current surveys and \textit{ATHENA}+.}
\label{area}
\end{figure}

\paragraph*{Active fraction and obscuration.}

We report the \textit{active} fraction $f_{\rm act}$ of SMBH progenitors, averaged over 30 simulations, in the labels of Figure \ref{density}. 
As it can be seen, $f_{\rm act}$ decreases with increasing redshift, from $f_{\rm act} = 37\%$ at $z = 7$ to $3\%$ at $z = 10$. 
On average, the total active fraction (at all redshifts) is $f_{\rm act} = 1.17 \%$. 
These values reflect the fact that BH growth is dominated by short, super-Eddington accreting episodes, particularly at high redshifts (P16), drastically reducing the fraction of active BHs, and thus the probability to observe them. 
A similar conclusion has been drawn by \citet{Page2001}, linking the observations of the local optical luminosity function of galaxies with the X-ray luminosity function of Seyfert 1. They find an active BH occupation fraction of $f_{\rm act} \sim 1 \%$. 
Comparable values have been also reported by \citet{Haggard2010} who combined \textit{Chandra} and SDSS data up to $z \sim 0.7$, and \citet{Silverman2009} for the 10k catalogue of the zCOSMOS survey up to $z \sim 1$.
While our predictions for $f_{\rm act}$ are consistent with the above studies, a larger fraction of active BHs is to be expected
in models where SMBH growth at $z > 6$ is Eddington-limited ($\sim 40 - 50 \%$ between $z \sim 7 - 10$, \citealt{Valiante2016}).

Figure \ref{fluxes} shows the total number of active progenitors as a function of flux in the \textit{Chandra} soft (0.5-2 keV) and hard (2-8 keV) bands.
We also distinguish super- (sub-) Eddington accreting BHs.
As a reference, we report the flux limits of \textit{Chandra} Deep Field South 4 Ms, $\rm F_{CDF-S}= 9.1 \times 10^{-18}$ $\rm erg \, s^{-1}\,cm^{-2}$ (dotted line, \citealt{Xue2011}) and \textit{Chandra} Deep Field North (CDF-N) 2 Ms, $\rm F_{CDF-N} = 2.5 \times 10^{-17}$ $\rm erg \, s^{-1}\,cm^{-2}$ (dot-dashed line, \citealt{Alexander2003}), showing for each panel and each band the average number N of active BHs with a flux larger than the limit of the CDF-S 4 Ms. 
In the upper panel we show the \textit{unabsorbed} model and the difference between the soft and hard X-ray band reflects the intrinsic SED. Moreover, since the flux limit of \textit{Chandra} is deeper in the soft band, this energy range is to be preferred for the detectability of high-z progenitors.

The effect of an isotropic absorption on the flux is shown in the bottom panel of Figure \ref{fluxes}. It does not appear to be as severe as it could be inferred from the large $N_{\rm H}$ shown in Figure \ref{propertymedd}. In fact, the soft (hard) \textit{Chandra} bands at $z = 7,8,9,10$ sample the rest frame energy bands $[4,16]\rm keV$, $[4.5,18]\rm keV$, $[5,20]\rm keV$, $[5.5,22]\rm keV$ ($[16,64]\rm keV$, $[18,72]\rm keV$, $[20,80]\rm keV$, $[22,88]\rm keV$), respectively. As discussed in Section \ref{abs}, in the range $[0.2-100]\rm keV$, the harder is the photon energy, the lower is the photoelectric absorption. 
As a result, the average number N of detectable BHs in the \textit{absorbed} model is close to that of \textit{unabsorbed} model at redshift $z \sim 7 - 8$, while it becomes much lower at larger $z$, reaching $\rm N = 0$ in the hard band at $z = 10$. This is a consequence of the larger fractions of Compton-thick BHs $f_{\rm CT}$ and, more generally, of the larger column densities. As already discussed, higher values of 
$N_{\rm H}$ correspond to super-Eddington accreting BHs. As a result, the shift towards lower fluxes in the \textit{absorbed} model mainly affects super-Eddington accreting BHs.

In the left panel of Fig.~\ref{numberflux} we show the cumulative number of BHs per unit area in the \textit{unabsorbed} (solid line) and \textit{absorbed} (dashed line) models with a flux $> F$ in the soft X-ray band.  We have assumed here an area of  $\rm \hat{A} = 0.03 \, \rm deg^2$ and show the flux limits of CDF-S 4 Ms and CDF-N 2 Ms as reference values\footnote{We assume BH progenitors to be distributed within a cube of 1 $\rm Gpc^3$, corresponding to an angular size of $A_{\rm box} \sim 390\times 390 \rm \, arcmin^{2}$ at $z \sim 7$ and $\sim 350 \times 350 \rm \, arcmin^2$ at $z \sim 10$.}.

For comparison, we report the number of AGN candidates selected with the same effective area coverage ($\rm A_{\rm obs} \sim \hat{A}$) by \citet{Giallongo2015} with a flux threshold of $\rm F_{\rm \hat{X}} = 1.5 \times 10^{-17} erg \, s^{-1} \, cm^{-2}$ (red circle). We also include the upper limit $\rm N < 1$ resulted from the analysis by \citet{Weigel2015} of the CDF-S.

In the \textit{unabsorbed} (\textit{absorbed}) model we find $\rm N(>F_{CDF-S}) = 0.15$ $(0.12)$ and $\rm N(>F_{\rm \hat{X}}) = 0.13$ $(0.1)$.
The effect of absorption decreases the number $\rm N$, also by a factor 2 for lower flux limits $(<-17)$, but it is not the main limiting factor preventing the observations of BH progenitors. In fact, we find that $N < 1$ also in the \textit{unabsorbed} model, for both $F_{CDF-S}$ and $F_{\rm \hat{X}}$. Our result is consistent with the non-detection reported by \citet{Weigel2015} and suggests that if the AGN candidates reported by \citet{Giallongo2015} are at $z > 6$, they are likely not SMBH progenitors of $z \sim 6$ quasars.
If we rescale linearly with $f_{\rm act}$ the relation in Figure \ref{numberflux}, for $f_{\rm act} = 1$ we would find an average number of observable active progenitors of $\rm N(>F_{CDF-S}) \sim 13$ $(10)$ and $\rm 
N(>F_{\rm \hat{X}}) \sim 11$ $(9)$. Thus, an active fraction of $f_{\rm act} < 10 \%$ is required in order to obtain a number of observed objects $N \lesssim 1$.

Interesting constraints on the activity of an early BH population have recently come from the measurement of the cross correlation signal between the fluctuations of the source-subtracted
cosmic infrared background (CIB) maps at 3.6 and 4.5 micron on angular scales $> 20''$ and the unresolved CXB at [0.5 - 2]~keV by \citet{cappelluti2013}. The authors argue that the
cross-power is of extragalactic origin, although it is not possible to determine if the signal is produced by a single population of sources (accreting BHs) or by different populations in the same area. 
Indeed, theoretical models show that highly obscured accreting black holes with mass $[10^4 - 10^6]~M_\odot$ at $z > 13$ provide a natural explanation for the observed signal 
\citep{Yue2013, yue2014}, requiring a number density of active BHs of $[2.7 - 4]\times10^{-5} \, M_\odot \, \rm Mpc^{-3}$ at $z \sim 13$ \citep{yue2016}.
While a detailed calculation of the cross-correlation between CXB and CIB is beyond the scope of the present analysis, in the right panel of Fig.~\ref{numberflux} 
we compare the CXB in the soft band predicted by our models with the upper limit of $3 \times10^{-13}/(1+z) \, \rm erg \,  cm^{-2} \, s^{-1} \, deg^{-2}$ placed by \citep{cappelluti2012} on the contribution of early black holes at $z > 7.5$ under the assumption that they produce the 
observed large scale CIB excess fluctuations \citep{kashlinsky2012}. For comparison, we also show the measured CXB in the soft band
reported by \citet{lehmer2012} from the analysis of the 4Ms  CDF-S.  The predictions for the absorbed and unabsorbed models
are more than a factor 10 below the upper limit by \citet{cappelluti2012}, showing that the cross-correlation signal can not be
reproduced by accreting SMBHs progenitors only.

\paragraph*{Best observational strategy.}

In order to understand which survey maximizes the probability to observe faint progenitors of $z \sim 6$ quasars, we define the number of BHs expected to be observed in a survey with sensitivity F and probing an area A of the sky:
\begin{equation}
N(F,A) = N(>F) \frac{A}{A_{\rm box}},
\end{equation}

\noindent where $N(>F)$ is the number of progenitors with flux $\geq F$.

In Figures \ref{area} we show $N(F,A)$ for the \textit{unabsorbed} (top panel) and \textit{absorbed} (bottom panel) models, in the \textit{observed} soft band. We report the contours corresponding to $N(F,A) = 10^{-2},10^{-1}$ (black dashed lines) and $N(F,A) = 1,10,10^2$ and $10^3$ (black solid lines).
For fluxes $\rm F_{[0.5-2]keV} \gtrsim 10^{-14} \, erg \, s^{-1}\,cm^{-2}$, we find $N(F,A) \lesssim 1$ for every possible area coverage.
We also show the sensitivity curves in the soft band of current surveys: CDF-S in yellow, \textit{AEGIS} in green \citep{Laird2009}, \textit{COSMOS Legacy} in cyan \citep{Civano2016}, XMM-LSS \citep{Gandhi2006} + XXL \citep{Pierre2016} in magenta. In white we show the predicted curve for \textit{ATHENA}+ with $5"$ PSF and multi-tiered survey strategy, for a total observing time of 25 Ms \citep[for details see][]{Aird2013}, and note that a survey can observe the integrated number $N(F,A)$ over its curve.
The difference between the \textit{unabsorbed} and the \textit{absorbed} models is almost negligible, reaching at most a factor of 2. In fact, the \textit{observed} soft-band corresponds, for high-z progenitors, to rest-frame energies hard enough to be almost unobscured, despite the large $N_{\rm H}$ and Compton-thick fraction (see Section \ref{results}).
The position occupied by the curve of the most sensitive survey performed nowadays, CDF-S, exploring a solid angle of $ 465 \rm \, arcmin^2$, is observationally disadvantaged with respect to the \textit{COSMOS Legacy}, less sensitive but covering a wider region of the sky.
This survey, in fact, should observe at least one progenitor. Similarly, XMM-LSS+XXL, despite having an even lower sensitivity, represent the current survey that maximizes the probability of SMBH progenitors detections.
A huge improvement in the detection will be obtained with \textit{ATHENA}+. According to our simulations, for a total observing time of 25 Ms more than 100 SMBH progenitors will be detected. 

The progenitors of $M_{\rm BH} \sim 10^9$ high-$z$ quasars are luminous enough to be detected in the X-ray soft band of current surveys. 
The real limit to their observability is that these objects are extremely rare, as a result of their low \textit{active} fraction. None of the surveys performed so far probes a region of the sky large enough for their detection to be 
meaningful, limiting the potentially observable systems to a few.

The above conclusion applies to a scenario where SMBH at z = 6 grow by short super-Eddington accretion episodes onto $100 M_\odot$
BH seeds formed at $z > 20$ as remnants of Pop~III stars. In \citet{Valiante2016} we have investigated the alternative scenario where BH
growth is Eddington limited and starts from BH seeds whose properties are set by their birth environment. According to this scenario, 
the formation of a few heavy seeds with mass $\sim 10^5 M_\odot$ (between 3 and 30 in our reference model) 
enables the Eddington-limited growth of SMBHs at $z > 6$. In a forthcoming paper, we will explore the X-ray detectability of SMBH progenitors
in this alternative scenario and make a detailed comparison with the results presented here.

\section{Conclusions}\label{conclusions}

The main aim of this work was to interpret the lack of detections of $z \gtrsim 6$ AGNs in the X-ray band.
Three are the most likely possibilities: \textit{i}) large gas obscuration, \textit{ii}) low BH occupation fraction or \textit{iii}) low \textit{active} fraction.

We developed a model for the emission of accreting BHs, taking into account the super-critical accretion process, which can be very common in high-z, gas-rich systems. We compute the spectrum of active BHs simulated by P16 with an improved version of the cosmological semi-analytical code \textsc{GAMETE/QSOdust}. In P16, we have investigated the importance of super-Eddington accretion in the early growth of $z \sim 6$ SMBHs. Here we model the emission spectrum of all the simulated SMBH progenitors at $z > 6$ and study their observability with current and future surveys.
Hence the sample of BHs that we have investigated does not necessarily represent a fair sample of \textit{all} BHs at $z > 6$ but only the sub-sample of those which contribute to the early build-up of the observed number of $z \sim 6$ quasars with mass $M_{\rm BH} \gtrsim 10^9 \, M_\odot$. 

We find that:
\begin{itemize}
\item the mean occupation fraction, averaged over 30 independent merger tree realizations and over the whole evolution, is $f_{\rm BH} = 35 \%$. It increases with time, being $f_{\rm BH} = 0.95,\, 0.84, \, 0.76,\,0.70$, at $z = 7, 8, 9, 10$, suggesting that the occupation fraction is not the main limitation for the observability of $z > 6$ BHs.

\item We find a mean Compton thick fraction of $f_{\rm CT} \sim 45\%$. Absorption mostly affect the super-Eddington accreting BHs at $z > 10$, where the surrounding gas reaches large values of $N_{\rm H}$;

\item Despite the large column densities, absorption does not significantly affect the \textit{observed} soft X-ray fluxes. In fact, at $z > 6$ the observed soft X-ray band samples the rest-frame hard energy band, where obscuration is less important.
The absorption can reduce the number of observed progenitors at most by a factor 2;

\item The main limiting factor to the observation of faint progenitors is a very low \textit{active} fraction, the mean value of which is $f_{\rm act} = 1.17 \%$. This is due to short, super-Eddington accreting episodes, particularly at high $z$. In fact, $f_{\rm act} = 3\%$ at $z = 10$ and grows to $f_{\rm act} = 37\%$ at $z = 7$ due to longer sub-Eddington accretion events.

As a result, surveys with larger fields at shallower sensitivities maximize the probability of detection. Our simulations suggest that the probability of detecting at least 1 SMBH progenitor at $z > 6$ is larger in the \textit{Cosmos Legacy} surveys than in the CDF-S. 
\end{itemize}

Better selection strategies of SMBH progenitors at $z > 6$ will be possible using future multi-wavelength searches. 
Large area surveys in the X-ray band (e.g. \textit{ATHENA+}) complemented with deep, high-sensitivity opt/IR observations (e.g. \textit{James Webb Space Telescope}) and radio detection may provide a powerful tool to study faint progenitors of $z \sim 6$ SMBHs.

\section{acknowledgements}

We thank Massimo Dotti, Enrico Piconcelli and Luca Zappacosta for their insightful help. We also acknowledge valuable discussions and suggestions from Angela Bongiorno, Marcella Brusa, Nico Cappelluti, Andrea Comastri, Roberto Gilli, Elisabeta Lusso and Francesca Senatore.
The research leading to these results has received funding from the European Research Council under the European 
Union’s Seventh Framework Programme (FP/2007-2013) / ERC Grant Agreement n. 306476.

\bibliography{Xray_cleaned}
\label{lastpage}

\end{document}